\documentclass[reprint,superscriptaddress,prl,amsmath,amssymb,aps]{revtex4-2}

\usepackage{graphicx}
\usepackage{dcolumn}
\usepackage{bm}
\usepackage{xcolor}
\usepackage{siunitx} 
\usepackage{relsize}
\usepackage{subfiles}
\usepackage{upgreek}
\usepackage{microtype} 
\definecolor{mycolor}{RGB}{0, 114, 189}
\usepackage[colorlinks=true, linkcolor=mycolor, citecolor=mycolor, urlcolor=mycolor, breaklinks=true]{hyperref}
\usepackage{physics}

\DeclareSIUnit\angstrom{\text {Å}}

\newcommand{\NV}{NV\textsuperscript{–}}
\newcommand{\NVs}{NV\textsuperscript{–}\hspace{-0.15em}s}

\usepackage{newfloat}
\DeclareFloatingEnvironment[
    name=Supplementary Fig.,
    fileext=lof,
    within=section,
    placement=tbhp]{extendedfigure}

\newcommand{\ATI}{Vienna Center for Quantum Science and Technology, Atominstitut, TU Wien, Stadionallee 2, A-1020 Vienna, Austria}
\newcommand{\TheoTU}{Institute for Theoretical Physics, TU Wien, Wiedner Hauptstraße 8-10, A-1040 Vienna, Austria}

\newcommand{\OIST}{Okinawa Institute of Science and Technology Graduate University, 1919-1 Tancha, Onna-son, Kunigami-gun, Okinawa 904-0495, Japan}

\begin{document}

\preprint{APS/123-QED}
\title{Self-Induced Superradiant Masing} 

\author{Wenzel Kersten}
\email{wenzelkersten@gmail.com}
\affiliation{\ATI}
\author{Nikolaus de Zordo}
\affiliation{\ATI}
\author{Oliver Diekmann}
\affiliation{\TheoTU}
\author{Elena S. Redchenko}
\affiliation{\ATI}
\author{Andrew N. Kanagin}
\affiliation{\ATI}
\author{Andreas Angerer}
\affiliation{\ATI}
\author{William J. Munro}
\affiliation{\OIST}
\author{Kae Nemoto}
\affiliation{\OIST}
\author{Igor E. Mazets}
\affiliation{\ATI}
\author{Stefan Rotter}
\affiliation{\TheoTU}
\author{Thomas Pohl}
\affiliation{\TheoTU}
\author{J{\"o}rg Schmiedmayer}
\email{schmiedmayer@atomchip.org}
\affiliation{\ATI}

\date{\today}

\begin{abstract}
    In cavity quantum electrodynamics (cQED) and particularly superradiance, emitters are typically assumed to be independent, interacting only through light shared via a common mode. While such photon-mediated interactions lead to a rich spectrum of collective optical effects, direct dipole-dipole interactions within the emitter ensemble are generally viewed as a source of decoherence. Here, we uncover a new role for direct spin-spin interactions as a drive for the superradiant dynamics of a hybrid system of nitrogen-vacancy center spins in diamond coupled to a superconducting microwave cavity. After an initial fast superradiant burst, we observe an unexpected train of subsequent emission pulses followed by quasi-continuous masing for up to one millisecond. We show that this surprising behavior arises from spectral hole refilling, where spin inversion is redistributed into the superradiant window of spins resonant with the cavity. We report measurements that clearly exclude other cQED-related effects, and performed microscopic simulations of up to one million spins, which demonstrate that the observed self-induced masing is indeed driven by dipole-dipole interactions between the spins. These findings open new pathways for exploring complex spin-spin interactions in dense disordered systems and offer possibilities for ultra-narrow linewidth solid-state superradiant masers powered purely by microwave-driven spin control.
\end{abstract}

\maketitle

Superradiance, first predicted by Dicke in 1954 \cite{dicke1954coherence}, describes an enhanced collective emission of light exhibiting high coherence and a nonlinear scaling of the intensity with the number of emitters \cite{gross1982superradiance,scully2009thesuper}. It has since been observed in many systems \cite{vanLoo1494,araujo2016superradiance,zhong2017interfacing,kim2018coherent,raino2018superfluorescence} including solid-state realizations with quantum dots \cite{scheibner2007superradiance,zhu2024single} and the negatively charged nitrogen-vacancy centers (\NV{}) in diamond \cite{bradac2017room,angerer2018superradiant}. This collective effect lies at the heart of cavity quantum electrodynamics \cite{walther2006cavity}, where emitters are typically treated as non-interacting, and their mutual couplings are mediated solely through the shared cavity field. These light-mediated interactions give rise to a variety of collective phenomena and phase transitions \cite{hepp1973superradiant,wang1973phase,baumann2010dicke,ferioli2023non}.

Building upon these fundamental explorations of collective light-matter interactions, recent advances have shifted the focus toward their applications in quantum technologies \cite{koppenhofer2022dissipative,bohr2024collectively}. 
Among other solid-state platforms \cite{gottscholl2022superradiance,wang2024tailoring,ng2024maser}, diamond-based systems stand out due to the exceptional quantum coherence and optical controllability of \NV{} centers \cite{bar2013solid}. Key examples include diamond-based microwave amplifiers \cite{sherman2022diamond}, quantum sensors \cite{wang2024spin}, mode cooling platforms \cite{macquarrie2017cooling,fahey2023steady}, and room-temperature diamond masers \cite{breeze2018continuous}.
Operating a diamond maser in a superradiant regime could further narrow its linewidth, with coherence maintained by the high cooperativity of the cavity-spin system itself \cite{bohnet2012steady}, rather than being dictated by the intra-cavity photon count as in conventional masers \cite{schawlow1958infrared}. This could pave the way for ultra-narrow linewidth sources, enabling high-precision frequency generation and quantum-limited microwave amplification. However, achieving a continuous-wave superradiant diamond maser remains an open challenge due to the need for both strong and uniform collective spin-photon coupling along with efficient optical pumping  \cite{wu2022superradiant}. While increasing the density of the spin ensemble enhances coupling, it also reduces optical transparency, complicating efficient optical pumping; moreover, it introduces greater dipole-dipole interactions, which can lead to decoherence~\cite{choi2017depolarization,zu2021emergent}. In this work, we show that such interactions can take on a more constructive role in dense spin systems like \NV{} centers. Our measurements reveal that dipole-dipole interactions can actively drive superradiant masing, which opens up a new self-induced regime of coherent collective emission in solid-state systems. We perform a comprehensive experimental characterization of the effect and conclusively trace it back to dipole-dipole interactions by numerically simulating the microscopic dynamics of up to one million \NV{} centers.

\begin{figure}
    \centering
    \includegraphics[width=85mm]{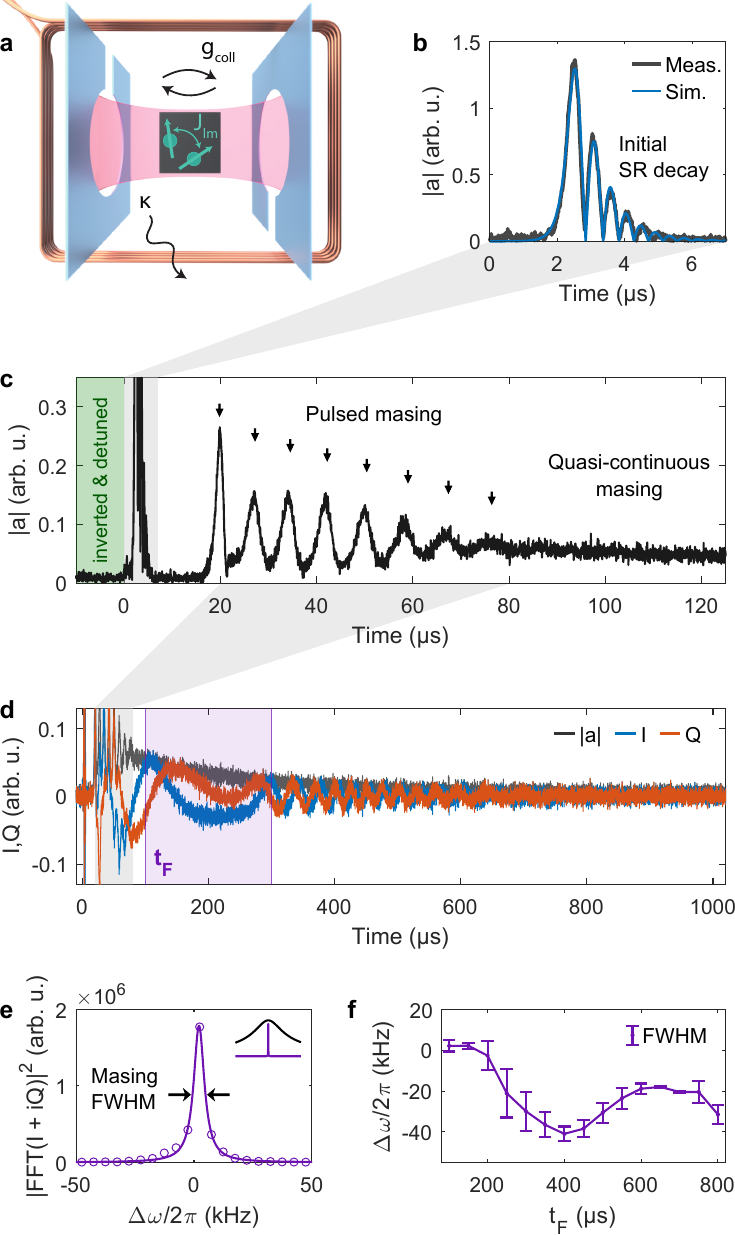}
    \caption{ \textbf{Experimental setup, self-induced superradiant masing dynamics and emission spectrum.}
    \textbf{a,} Schematic of the superconducting microwave cavity strongly coupled to the \NV{} diamond.
    \textbf{b,} Zoomed-in plot of the cavity amplitude $|a|$ during the initial superradiant (SR) decay, which is triggered upon tuning the inverted spin ensemble back into resonance and is well described by a semiclassical model. 
    \textbf{c,} Expanding the time-axis after the initial superradiant decay, we observe a series of narrow masing pulses evolving into a quasi-continuous cavity emission. Note the different $y$-axis scalings for $|a|$ as compared to panel \textbf{b}.
    \textbf{d,} Long tail of the quasi-continuous masing emission, showing the quadratures $I$ and $Q$ of the cavity amplitude, digitally demodulated in the rotating frame of the cavity resonance frequency for visual clarity. The purple shaded area marks the interval for the Fourier analysis, which is plotted in 
    \textbf{e,} where the emission has a linewidth much smaller than the cavity, see inset. The frequency difference $\Delta \omega /2\pi$ is measured relative to the cavity frequency of $\SI{3.1}{\giga \hertz}$. 
    \textbf{f,} The frequency and linewidth of the emission changes over time when the window of the Fourier analysis is shifted (see text for details).} 
    \label{fig1:setup}
\end{figure}

In our experiment, we investigate an inhomogeneously broadened ensemble of \NV{} centers strongly coupled to a microwave cavity. Our system, illustrated in Fig.~\ref{fig1:setup}a, allows for the generation of an inverted spin ensemble by applying strong microwave pulses, and for the controlled release of this inversion in the form of superradiant emission. The cavity, made of two parallel sapphire chips with superconducting split-ring structures, has a resonance frequency of around $\omega_c/ 2 \pi = \SI{3.1}{\giga \hertz}$. The roughly $\SI{200}{\micro \meter}$ sized diamond sample is positioned between the chips. A wire loop wrapped directly around the chips is used for rapid switching of the external magnetic field, allowing us to activate or suppress the cavity-spin interaction on demand. This assembly is housed in a copper box inside a dilution refrigerator, cooled to $\SI{25}{\milli \kelvin}$ and connected to a heterodyne detection scheme. The hybrid system of cavity and spin ensemble, with $N \approx \num{9e12}$ \NV{} centers, is well within the collective strong coupling regime, having a cooperativity $C = 15.2$. Uniform cavity-spin coupling over the sample volume guarantees the required spin permutation symmetry to enable the superradiant dynamics we observe. In this dense spin ensemble, the typical nearest-neighbor distance between \NV{} spins is $r =(N/V)^{-1/3}\approx \SI{8}{\nano \meter}$. This results in a typical spin-spin coupling strength of about $\SI{100}{\kilo \hertz}$ between neighboring \NVs{}. Further details on the diamond sample, cavity, and setup are provided in the Supplementary Material.

\vspace{0.4cm} \noindent \textbf{\large Results} \vspace{0.2cm} \\ \noindent
\noindent \textbf{Initial superradiant decay and spectral hole} \vspace{0.2cm} \\ \noindent
For all experiments, we employ the previously established protocol \cite{kersten2023triggered} for generating a uniformly inverted spin state. All \NV{} spins are tuned into resonance with the cavity, using a static magnetic field with equal projections along the four diamond axes. We apply a microwave inversion pulse to homogeneously invert all spins from a relaxed initial state of the effective two-level systems. Subsequently, we rapidly detune the spin ensemble from the cavity resonance and store the inversion for a set hold time. This procedure allows for the preparation of states with uniform initial spin inversion $p_0 = \langle \sigma_j^z \rangle$ and almost zero transversal spin components $\langle \sigma_j^- \rangle \approx 0$. The values for $p_0$, bounded by $\pm 1$, are tunable within the range of $0.1$ to $0.4$ by modifying the hold time on the order of milliseconds.

Upon tuning the spins back into resonance with the cavity using the detuning loop, the inverted spin state is free to interact with the cavity mode. If the stored inversion exceeds the threshold $p_0C > 1$, the system enters a metastable state~\cite{molmer1}. Here, $C = g_\mathrm{coll}^2/\kappa \Gamma \approx 15.2$ is the cooperativity, a dimensionless parameter combining the collective coupling strength $g_\mathrm{coll}/2\pi = \SI{4.53}{\mega \hertz}$, the cavity linewidth $\kappa/2 \pi = \SI{418}{\kilo \hertz}$ (half width at half maximum), and the effective ensemble dephasing rate $\Gamma / 2 \pi = \SI{3.22}{\mega \hertz}$. The latter accounts for both the inhomogeneously broadened spin frequency distribution $\rho(\omega)$, with a full width at half maximum (FWHM) of $W/2\pi = \SI{8.65}{\mega \hertz}$, and the spin linewidths modeled by $\gamma_\perp/2\pi = \SI{179}{\kilo \hertz}$, see Eqs.~(\ref{eq:CDelta}) and (\ref{eq:total_coop}) in the Supplementary Material.

In this metastable inverted state with $p_0C > 1$, any fluctuation will stimulate a collective emission process known as a superradiant decay \cite{skribanowitz1973observation}. Conversely, if the inversion is below the instability threshold, dephasing due to inhomogeneous broadening becomes dominant and prevents this avalanche process \cite{molmer1}. In our case, the superradiant decay is triggered by noise photons from the input line. As the spin decay accelerates, the cavity amplitude $|a|$ increases. 
It reaches its maximum as the collective spin vector points towards the equator of the Bloch sphere, where the cavity amplitude $\max(|a|) \propto p_0 - 1/C$ serves as a measure of the initial inversion above threshold, see Eq.~(\ref{eq:max_a_vs_p}) in the Supplementary Material, and the emitted intensity $|a|^2 \sim (p_0N)^2$ exhibits the characteristic quadratic scaling of superradiance with the number of effectively participating spins \cite{kersten2023triggered, gross1982superradiance}. Subsequently, energy is coherently exchanged between cavity and spins, visible as damped Rabi oscillations in Fig.~\ref{fig1:setup}b. This coherent exchange is eventually stopped by coherence-limiting processes in the system, mainly the dephasing of the inhomogeneously broadened spins. 

Crucially, the cavity-resonant spins --- which dominate the collective emission dynamics --- become de-excited through the superradiant emission, while the off-resonant spins maintain their inversion. This leaves behind a \emph{spectral hole} \cite{putz2017spectral}, a region of depleted inversion centered at the cavity frequency. This initial superradiant decay dynamics is well-captured by the Maxwell-Bloch equations \cite{carmichael1999statistical} that provide a semiclassical description of the collective coupling between a non-interacting spin ensemble and the cavity mode. 
This standard picture does not predict further dynamics beyond the initial emission pulse, once the resonant spin inversion has been sufficiently depleted below the threshold $p(\Delta\,{=}\,0) < 1/C$, see Eq.~(\ref{eq:weighted_instability}).

\vspace{0.15cm} \noindent \textbf{Pulsed and quasi-continuous masing} \vspace{0.2cm} \\ \noindent
Surprisingly, however, we observe a train of masing pulses that emerges about $\Delta t \approx \SI{15}{\micro \second}$ after the initial superradiant burst, as shown in Fig.~\ref{fig1:setup}c. This behavior cannot be understood within the standard semiclassical theory for emitter ensembles in cavities \cite{carmichael1999statistical} and suggests a dynamical refilling of the spectral hole burned by the collective cavity emission. 
The timescale for the revival pulses appears unexpectedly long, greatly exceeding the characteristic timescales of the cavity loss rate $\kappa^{-1}$, the effective ensemble dephasing $\Gamma^{-1}$, the collective cavity coupling $g_\mathrm{coll}^{-1}$, and the spin decoherence $\gamma_\perp^{-1}$. This clearly excludes Rabi oscillations or spin-echo effects as potential explanations~\cite{weichselbaumer2020echo} for the observed superradiant pulse trains.

The masing pulses appear as distinct Gaussian-like peaks in the cavity amplitude with progressively increasing full widths at half maximum (FWHM) ranging from around $\SI{1.5}{\micro \second}$ to $\SI{3.8}{\micro \second}$. Each pulse has a random but nearly constant phase as determined from the IQ-quadratures of the cavity amplitude, corresponding to coherent pulsed masing with near transform-limited FWHM bandwidths ranging from around $\SI{400}{\kilo \hertz}$ to $\SI{120}{\kilo \hertz}$. Following the early sequence of discrete pulses, the emission evolves into a quasi-continuous regime that persists for up to $\SI{1}{\milli \second}$.

The extended cavity dynamics, shown in Fig.~\ref{fig1:setup}d, is measured with the same system parameters as in Fig.~\ref{fig1:setup}c but at half the digitizer sample rate in the heterodyne detection chain. The recorded quadratures $I$ and $Q$ of the cavity amplitude are demodulated at an intermediate frequency of $\SI{5}{\mega \hertz}$ detuned from the cavity resonance. We extract the emission's linewidth by employing fast Fourier transform (FFT) analysis and Lorentzian profile fitting within the integration window of $\SI{200}{\micro \second}$ (see Fig.~\ref{fig1:setup}e). Shifting the start of the integration window $t_\text{F}$, we analyze in Fig.~\ref{fig1:setup}f how the central frequency of the masing emission drifts over time within $\pm \SI{25}{\kilo \hertz}$. The observed linewidth, varying from $\SI{5}{\kilo \hertz}$ to $\SI{20}{\kilo \hertz}$, is two orders of magnitude below the cavity linewidth $\kappa$ and the individual spin linewidth $\gamma_\perp$, highlighting the crucial role of collective enhancement --- an indicative trait of superradiance --- in achieving high coherence. We attribute the linewidth variation to the emission frequency drift within the integration window caused by magnetic field oscillations after rapidly switching the detuning loop. These oscillations appear on a scale that is three orders of magnitude smaller than the full extent of the loop detuning of roughly $\SI{20}{\mega \hertz}$.

\begin{figure}[t]
    \includegraphics[width=85mm]{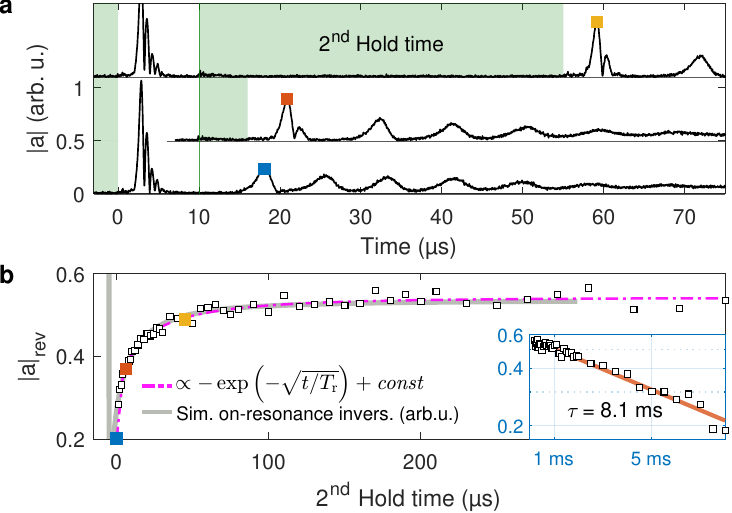}
    \caption{\textbf{Influence of second hold time on revival dynamics.}
    \textbf{a,} Stacked cavity signals of the superradiant dynamics with a second stabilization sequence (hold time), represented by light green shading. Increasing the duration of this second hold time extends the spectral hole-filling process, influencing the amplitude of the superradiant masing pulse revival. 
    \textbf{b,} Revival amplitude $|a|_\mathrm{rev}$ for varying second hold times. An initial stretched exponential increase is followed by an exponential decrease for longer timescales (see inset). The data points are well described by simulating the on-resonance inversion using the parameters of Fig.~\ref{fig3:spectralholesim}.}
    \label{fig2:secondholdtime}
\end{figure}

\begin{figure*}[t]
    \includegraphics[width=180mm]{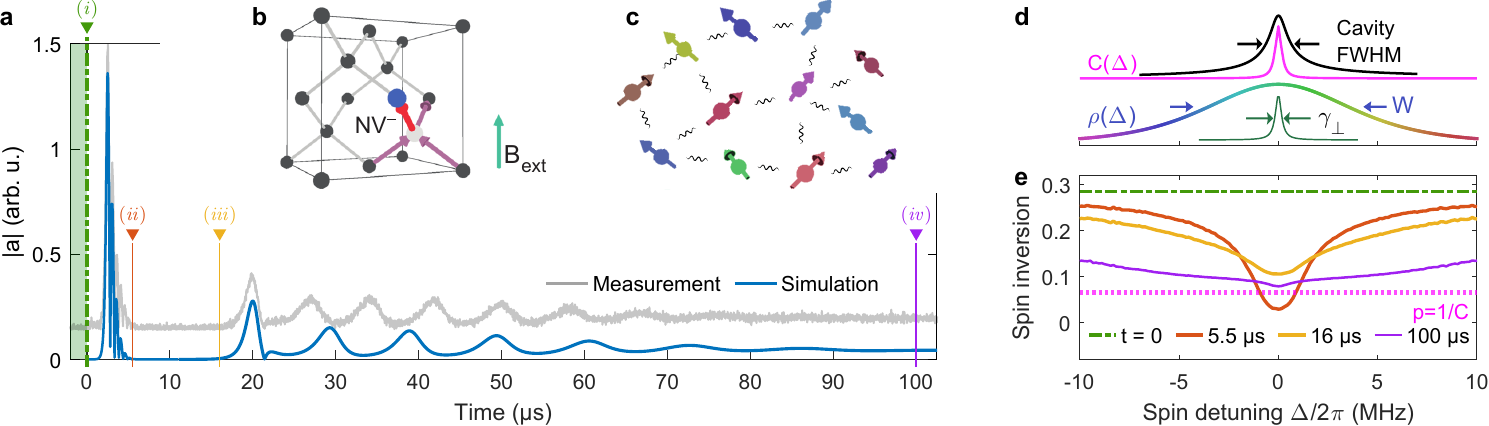}
    \caption{
    \textbf{Simulation of superradiant dynamics driven by spin-spin interactions.}
    \textbf{a,} Simulation of the cavity amplitude $|a|$ based on the microscopic spin model compared to measurement (offset vertically for clarity), with four key timesteps marked (cf. \textbf{e}).
    \textbf{b,} \NV{} center in the diamond unit cell, illustrating one of four possible alignments.
    \textbf{c,} Schematic of the simulated spin network: \NV{} centers are randomly distributed in space, orientation, and resonance frequency, and interact via dipole-dipole couplings.
    \textbf{d,} Relevant linewidths in the hybrid system: cavity linewidth, inhomogeneous spin distribution of width $W$, single-spin dephasing rate $\gamma_\perp$, and the frequency-dependent cooperativity function $C(\Delta)$.
    \textbf{e,} Simulated spin inversion profiles at four key times: (\textit{i}) at $t=0$: uniform inversion, (\textit{ii}) at $t=\SI{5.5}{\micro \second}$: deep spectral hole, (\textit{iii}) at $t=\SI{16}{\micro \second}$: refilled above threshold, (\textit{iv}) at $t=\SI{100}{\micro \second}$: broad, depleted quasi-steady state.}
    \label{fig3:spectralholesim}
\end{figure*}

\vspace{0.15cm} \noindent \textbf{Experimental evidence for spectral hole refilling} \vspace{0.2cm} \\ \noindent
To identify the mechanism responsible for the apparent spectral hole refilling, we conduct a set of auxiliary measurements designed to decouple the dynamics within the spin system from the cavity. Right after the initial superradiant decay, we rapidly detune the spins and introduce a second hold time, during which spin-cavity interaction is suppressed. Extending this isolation phase reveals that the amplitude of the first revival pulse increases with longer off-resonant hold times, as shown in Fig.~\ref{fig2:secondholdtime}a,b. This strong dependence on the isolation period clearly establishes that the cavity coupling plays no significant role in the spectral hole refilling, which instead must be driven by another additional mechanism. When the spins are tuned back into resonance, the accumulated inversion inside the hole --- now above the instability threshold --- triggers a superradiant pulse, whose amplitude scales with the amount of resonant spin inversion, which initially increases with the hold time. At longer hold times, once the spectral hole is maximally refilled, the amplitude reaches a plateau, as shown in Fig.~\ref{fig2:secondholdtime}b. The eventual decrease in amplitude over longer timescales reflects a global loss of spin inversion that has been previously discussed in Ref.~\cite{kersten2023triggered}.

Additionally, in Supplementary Fig.~\ref{figSup:holeburning}, we closely reproduce the revival pulse dynamics by deliberately creating a spectral hole with a microwave pulse, confirming that the refilling mechanism operates independently of how the hole is initially formed. 
While the observed pulsed masing bears resemblance to recently reported periodic superradiance in optically pumped spin systems \cite{hara2024periodic}, our system operates in a fundamentally different regime: the pulses emerge spontaneously without any external pump, pointing to a self-induced mechanism originating within the spin ensemble.

\vspace{0.15cm} \noindent \textbf{Many-body dynamics through dipole-dipole-interactions in an \NV{} ensemble} \vspace{0.2cm} \\ \noindent
In order to understand and explain our experimental observations, we have performed microscopic simulations of the cavity-coupled spin ensemble in the presence of direct interactions between the emitters. As detailed in the Supplementary Information, the magnetic dipole interactions between the \NV{} centers generate a coherent exchange of spin excitations, which offers a possible mechanism for transporting inversion across the energy spectrum of the inhomogeneously broadened spin ensemble. 
The proper description of the collective spin-cavity coupling and the resulting superradiant emission dynamics requires sizable particle numbers in order to reliably sample the random positions, frequencies, and orientations of the spins. Hence, we have implemented microscopic simulations of the resulting many-body dynamics for large numbers of $\sim10^6$ particles, with randomly sampled positions ${\bf r}_j$ in a cubic box to resemble our experimental \NV{} density, random cavity detunings $\Delta_j$ drawn from the known experimental spectrum $\rho(\Delta)$ (cf. Fig.~\ref{fig3:spectralholesim}d), and randomly chosen spin orientations from one of the four possible directions for the tetrahedral symmetry of diamond as illustrated in Fig.~\ref{fig3:spectralholesim}b. Noting that the single-spin decoherence rate $\gamma_\perp$ exceeds the typical strength $J_{jk}$ of pairwise dipole-dipole interactions, one can derive an effective rate description for the interaction-driven evolution of the spin inversion
\begin{align}
    \partial_t{p_j} = -\sum_k \frac{4\gamma_\perp |J_{jk}|^2}{(\Delta_j-\Delta_k)^2+4\gamma_\perp^2}\left(p_j-p_k\right),
    \label{eq:diffusion}
\end{align}
which in turn can strongly affect the many-body emission dynamics due to the collective cavity coupling. Equation~(\ref{eq:diffusion}) describes the transport of spin excitations across the sample and across the energy spectrum of the spin ensemble with a two-body rate $|J_{jk}|^2 \propto 1/|{\bf r}_k-{\bf r}_j|^6$ that is only effective for closely adjacent \NV{} centers. 

In Fig.~\ref{fig3:spectralholesim}, we show the simulation results for the spin-cavity dynamics in the presence of dipole-dipole interactions. We infer a starting inversion of $p_0=0.285$ from the initial superradiant decay and set the typical nearest-neighbor \NV{} distance to $r \approx \SI{7}{\nano \meter}$ to align the onset of the pulsed masing with the experimental data, in good agreement with the independently estimated value of $\SI{8}{\nano\meter}$ based on the collective coupling strength and cavity mode volume (see Supplementary Material). 

The simulation quantitatively reproduces the dynamics up to the first revival pulse and qualitatively captures the key features of the superradiant emission throughout the entire measurement interval, from the initial sequence of periodic pulses followed by the quasi-continuous masing regime. While the precise spacing of the periodic pulses is not perfectly reproduced, the theory captures the characteristic shape of the masing pulses as well as their steadily increasing delay times. 

To further corroborate the important role of dipole-dipole interactions, we also studied the refilling of the spectral hole in the absence of spin-cavity coupling, as measured experimentally (Fig.~\ref{fig2:secondholdtime}). We use the fact that the amplitude of a superradiant pulse directly results from the cavity-resonant spin inversion just before pulse onset, cf. Eq.~(\ref{eq:max_a_vs_p}). The measurement of $|a|_\mathrm{rev}$ (Fig.~\ref{fig2:secondholdtime}a) thus provides direct access to the evolution of $p(\Delta\,{=}\,0)$, and is shown as a function of the hold time in Fig.~\ref{fig2:secondholdtime}b. For comparison we have simulated the spin dynamics, using the same parameters as in Fig.~\ref{fig3:spectralholesim}, but switching off the spin-cavity interaction after the initial superradiant decay. After accounting for a time offset between the initial decay and the first revival pulse, we find excellent agreement between experiment and theory, and observe that the hole-refilling dynamics follow a stretched exponential
\begin{equation}
    p(\Delta=0)-\overline{p}\propto \exp(-\sqrt{t/T_\mathrm{r}}),
    \label{eq:strechedExp}
\end{equation}
where $\overline{p}$ is the average inversion of the ensemble. 
Fitting the revival pulse amplitude as a function of the second hold time yields a characteristic time constant $T_\mathrm{r}\approx \SI{11.6}{\micro \second}$, as shown in Fig.~\ref{fig2:secondholdtime}b. 
This behavior is consistent with an independent measurement of the refilling time, obtained from the time delay $\Delta t$ between the initial superradiant decay and the first revival pulse for different initial inversions $p_0$, see Supplementary Fig.~\ref{figSup:hole_depth} for details.
The stretched exponential relaxation emerges from a simplified analytical solution to Eq.~(\ref{eq:diffusion}), as detailed in the Supplementary Information. Importantly, the observed exponent of $1/2$ (see Supplementary Fig.~\ref{figSup:stretchexponent}) is a hallmark of the $1/r^3$ dependence of dipolar interactions~\cite{choi2017depolarization}. We thus conclude that dipole-dipole interactions between the spins are indeed responsible for the transfer of inversion with and without spin-cavity interactions. 

Based on the above insights, we can now draw a conclusive picture of the observed superradiant masing dynamics. To this end, we show in  Fig.~\ref{fig3:spectralholesim}e the inversion profile at four key times during the measured and simulated dynamics: (\textit{i}) All spins start fully inverted, well above threshold. (\textit{ii}) The initial superradiant decay generates a deep spectral hole around the cavity resonance, lowering the spin inversion below threshold. Now, dipole-dipole interactions steadily repopulate the on-resonance spin inversion $p(\Delta\,{=}\,0)$. (\textit{iii}) The first revival pulse emerges after the refilled inversion crosses the instability threshold, triggered by residual cavity photons. Its amplitude reflects how much $p(\Delta\,{=}\,0)$ has grown above threshold. Following the simultaneous creation of the spectral hole, the refilling process starts anew. As more and more energy is emitted by the cavity, each subsequent pulse has fewer participating excited spins and thus generates a shallower hole, which leads to faster refilling. At the same time, the superradiant build‐up gradually slows down, due to the $1/p$ scaling of the associated timescale \cite{RAMI}. 
As a result, the system transitions from an initial pulsed regime to a quasi-steady state (\textit{iv}) where energy is continuously replenished from the off-resonant spins, while simultaneously being emitted through the cavity. This process eventually halts when the masing cannot be upheld due to significant inversion loss.

\vspace{0.4cm} \noindent \textbf{\large Discussion} \vspace{0.2cm} \\ \noindent
We have demonstrated a novel manifestation of superradiant masing in a hybrid system composed of nitrogen-vacancy center spins in a diamond coupled to a superconducting microwave cavity. Our observation of self-induced pulsed and quasi-continuous superradiant masing is well reproduced by a microscopic model that implements dipole-dipole interactions together with the coupling to the cavity. Remarkably, dipolar interactions, far from being merely a decoherence channel, result in a redistribution of energy that enables collective emission in a strongly disordered ensemble.

This behavior highlights that dense dipolar spin networks, often idealized as formally closed many-body systems \cite{yao2014many,nandkishore2015many}, can effectively act as their own bath and relax toward thermal-like states on microsecond timescales. A finite $T_2 = 1/\gamma_\perp$  (combining intrinsic and background-spin-induced broadening) ensures sufficient spectral overlap and enforces equilibration. An open question is whether this self-thermalization --- and the associated hole refilling --- can be suppressed or modified by ensemble-splitting or dynamical-decoupling protocols, a subject for future exploration on our tunable \NV{} platform.

Beyond fundamental insights, our findings suggest a route to a solid-state superradiant maser driven by spin-spin interactions. By repeatedly re-exciting off-resonant spins, one could sustain inversion without optical pumping \cite{breeze2018continuous}. This concept parallels recent diamond amplifiers that exploit cross-relaxation between \NV{} centers and nitrogen impurities \cite{ohta2025near}, and may enable ultra-narrow-linewidth microwave sources for precision metrology \cite{wu2022superradiant}.

\clearpage
\vspace{0.15cm} \noindent \normalsize \textbf{Acknowledgments}\\ \footnotesize
We thank E. Demler, H. Ritsch, S. Yelin, and T. Zhang for fruitful discussions. We acknowledge support by the Austrian Science Fund (FWF) projects I3765 (MICROSENS), P34314 (Spins in Quantum Solids). W.J.M. and K.N. would like to acknowledge support from the MEXT Quantum Leap Flagship Program (MEXT QLEAP) Grant No. JPMXS0118069605.

\vspace{0.15cm} \noindent \normalsize \textbf{Author contributions} \\ \footnotesize
W.K. designed the cavity, worked on the experimental setup, performed the measurements together with N.d.Z., and carried out the data analysis together with O.D., who, together with A.A., I.E.M., S.R., and T.P. developed the microscopic theory. A.N.K. contributed to building the measurement setup. The experiments were conceived and conceptualized by W.K., W.J.M., and K.N. The manuscript was written by W.K., N.d.Z., O.D., and E.S.R., with input from all authors. J.S. and E.S.R. supervised the project.

\bibliography{Self-Induced_Superradiant_Masing}


\renewcommand{\theequation}{S\arabic{equation}}
\setcounter{equation}{0}

\onecolumngrid
\newpage
\normalsize
\appendix
\begin{center}
\textbf{\Large Supplementary Material for: Self-Induced Superradiant Masing}
\end{center}

\vspace{0.5cm}
The Supplementary Material is structured in three main parts: (\textit{i}) \textbf{Supplementary Figures}, containing extended data and supporting plots; (\textit{ii}) \textbf{Methods}, providing additional details on the experimental procedures and theoretical modeling; and (\textit{iii}) \textbf{Theory Supplement}, presenting in detail the analytical framework underlying the microscopic simulations, including a discussion of NV orientations, the mean-field description, and the origin of the stretched exponential relaxation behavior.

\begin{center}
\rule{4cm}{0.5pt}
\end{center}
\vspace{2cm}

\begin{center}
\textbf{\large Supplementary Figures}
\end{center}

\begin{extendedfigure}[b]
    \includegraphics[width=90mm]{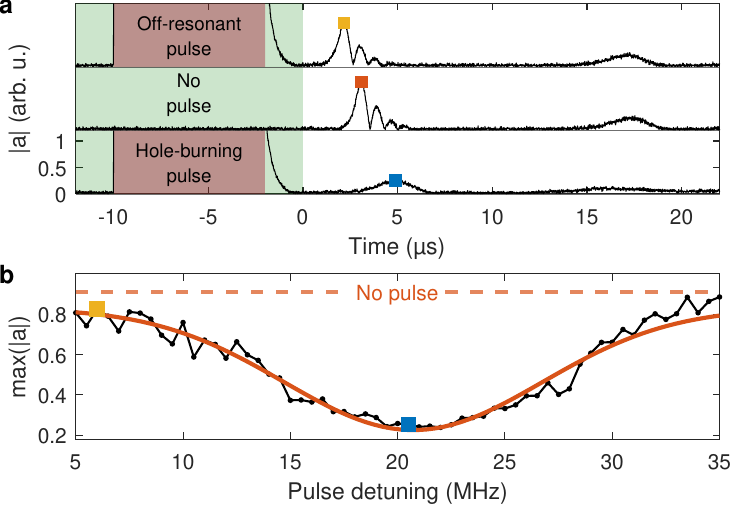}
    \caption{
    \textbf{Probing superradiant dynamics after a hole-burning pulse.}
    \textbf{a,} Cavity dynamics when applying a strong microwave pulse (red shading) to the spins while detuned (green shading) and subsequently triggering the superradiant decay and revival pulses. This will create a spectral hole in the ensemble, resulting in a slice of decreased spin inversion at varying frequencies for different runs. A resonant hole-burning pulse reproduces a spectral hole similar to the one created by the initial superradiant decay when no hole-burning pulse is applied. When the hole-burning pulse is off-resonant at the sides of the spin distribution, it still leads to an earlier triggering of the initial superradiant decay, but will otherwise recreate similar dynamics as when no pulse is applied. \textbf{b,} Maximum amplitude of the superradiant decay pulse for different hole-burning frequencies, revealing the frequency distribution of the detuned spins, although appearing with a larger width due to power broadening.}
    \label{figSup:holeburning}
\end{extendedfigure}

\clearpage
\newpage

\begin{extendedfigure}[t]
    \includegraphics[width=180mm]{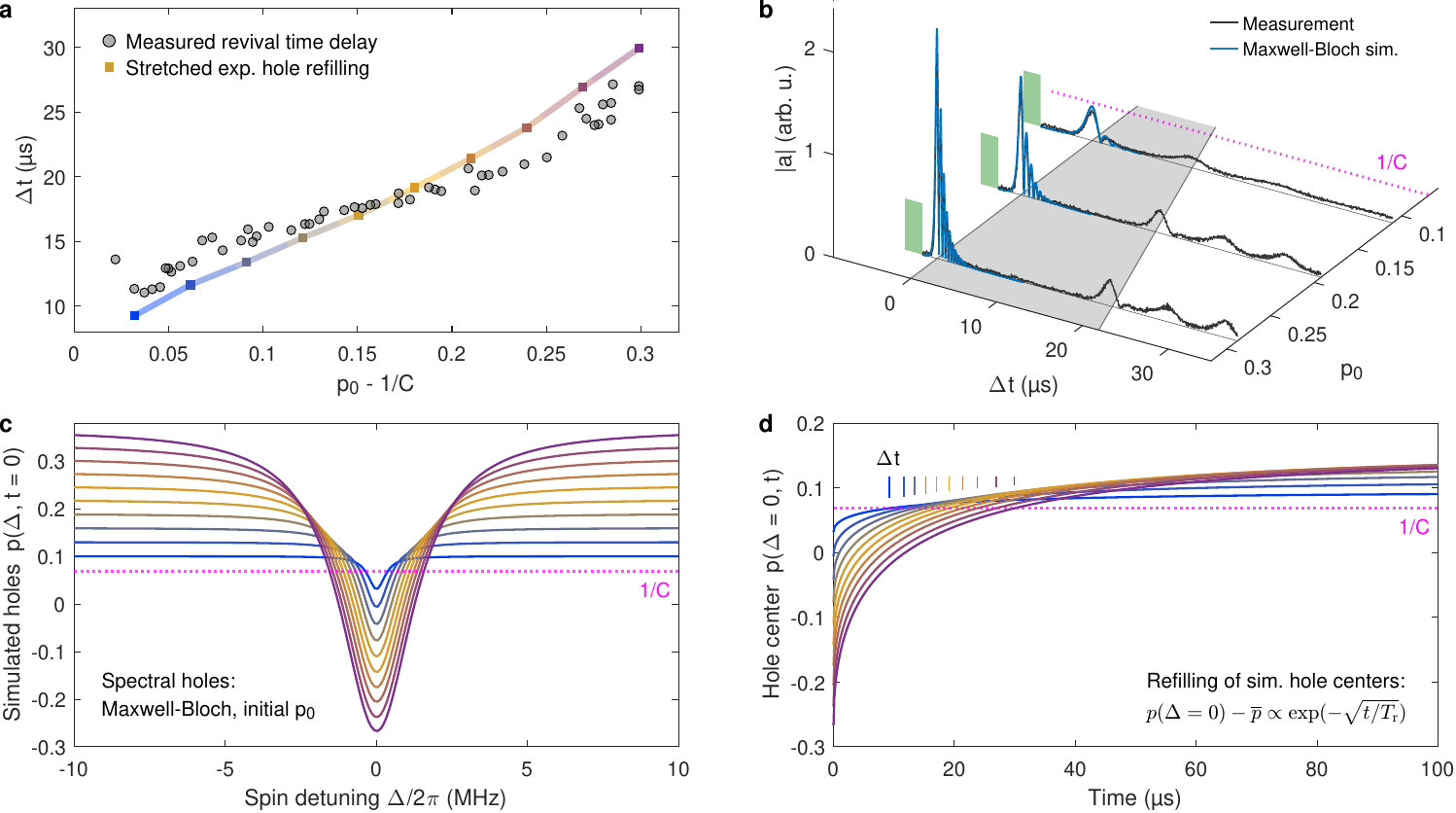}
    \caption{
    \textbf{Influence of initial inversion on refilling time.}    
    \textbf{a,} Measured time delay $\Delta t$ between the initial superradiant decay and the first revival pulse (circles), plotted as a function of the reduced initial inversion $p_0 - 1/C$. This reduced inversion corresponds to the depth of the spectral hole created by the initial emission, since the resonant spins involved effectively undergo a $\pi$-rotation (cf. panel \textbf{c}). 
    \textbf{b,} Experimental runs for three different initial inversion levels $p_0$, illustrating how $\Delta t$ is extracted from the cavity signal $|a|$.
    The values of $p_0$ are obtained by fitting the initial superradiant decay using Maxwell-Bloch equations [Eqs.~(\ref{eq:MBE_all}) without spin-spin interactions]. 
    \textbf{c,} From such Maxwell-Bloch simulations, we obtain access to the inversion profile. These profiles are shown for different initial $p_0$, evaluated at the time when the cavity amplitude $|a|$ reaches its peak value during the initial superradiant pulse. In all cases, the initial inversion $p_0$ is assumed to be uniform across the spin ensemble.
    \textbf{d,} Using the stretched exponential relaxation of the on-resonance inversion, Eq.~\eqref{eq:strechedExp}, we can estimate the refilling dynamics with the experimentally obtained time constant $T_\mathrm{r} = \SI{11.6}{\micro\second}$ from the fit of Fig.~\ref{fig2:secondholdtime}b. We extract from these curves the time $\Delta t$ when the on-resonance inversion reaches $1/C$ (squares in panel \textbf{a}). This time delay $\Delta t$ serves as a measure for how long it takes to refill the spectral hole above the instability threshold $p(\Delta\,{=}\,0) > 1/C$, enabling the revival pulse. We find good agreement between the experiment and the estimated time delay.}
    \label{figSup:hole_depth}
\end{extendedfigure}

\begin{extendedfigure}[b]
    \includegraphics[width=180mm]{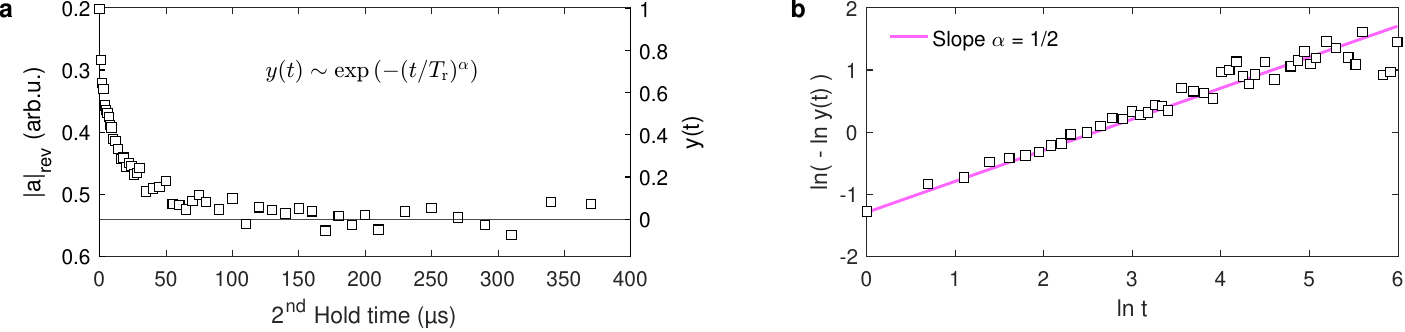}
    \caption{
    \textbf{Stretching exponent evaluation of spectral hole refilling.}  
    \textbf{a,} Data from Fig.~\ref{fig2:secondholdtime}b, normalized to define the relaxation function $y(t)$, which decays from 1 to 0 with a characteristic stretching exponent $\alpha$.  
    \textbf{b,} Double-logarithmic plot of $\ln(-\ln y(t))$ versus $\ln t$. The linear trend confirms stretched exponential dynamics, and the slope yields a direct estimate of the stretching exponent $\alpha \approx 1/2$. This value is direct evidence of the expected $1/r^3$ scaling of dipolar spin-spin interactions in three dimensions, see Supplementary Information and Ref.~\cite{choi2017depolarization}.}
    \label{figSup:stretchexponent}
\end{extendedfigure}

\clearpage
\newpage

\begin{extendedfigure}[t]
    \includegraphics[width=180mm]{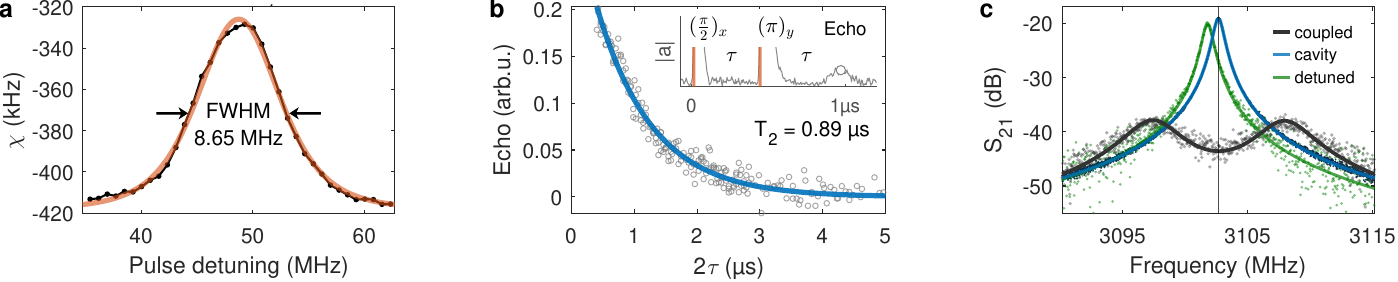}
    \caption{
    \textbf{Spin distribution, $T_2$, and steady-state spectroscopy measurements.}
    \textbf{a,} Measurement of the inhomogeneously broadened spin distribution in the dispersive regime, with spins detuned by around $\delta/2\pi \approx \SI{50}{\mega \hertz}$ from the cavity. Following a $\SI{100}{\milli \second}$ weak microwave drive at different frequencies, with the spin system initially in the ground state, only spins with the corresponding frequency will end up in an equal mixture of up and down spins, while other spins are unaffected. This change in the overall spin ensemble inversion leads to a dispersive shift $\chi \propto \langle S_z \rangle$ of the cavity resonance frequency, measured using a vector network analyzer.
    \textbf{b,} Measurement of $T_2$ using a Hahn echo sequence with $(\frac{\pi}{2})_x$ and $(\pi)_y$ pulses of $\SI{20}{\nano \second}$ duration.
    \textbf{c,} Transmission spectroscopy of the hybrid system using high (blue) and low (black) microwave power. Using high input power, the whole spin ensemble is brought into a completely mixed state, effectively decoupling it from the cavity. A Lorentzian fit yields a linewidth of $\kappa/2\pi = \SI{418}{\kilo \hertz}$. Using low input power, the normal-mode splitting of the coupled system in the ground state is measured and fitted with a collective coupling strength of $g_\mathrm{coll}/2\pi = \SI{4.53}{\mega \hertz}$. When the spins are detuned (green), the response at low input power shows the suppression of cavity-spin interaction.}
    \label{figSup:systemcharact}
\end{extendedfigure}

\begin{center}
\rule{4cm}{0.5pt}
\end{center}
\vspace{1cm}

\begin{center}
\textbf{\large Methods}
\end{center}
\vspace{0.4cm} \noindent \textbf{Spin ensemble} \vspace{0.2cm} \\ \noindent
The spin system used in this work is the negatively charged nitrogen-vacancy center in diamond. The ground state Hamiltonian of the defect with spin $S=1$ is given by $\mathcal{H} = \hbar D S_z^2 + \mu \mathbf{B} \cdot \mathbf{S}$, with the zero field splitting $D/2 \pi = \SI{2.88}{\giga \hertz}$ and $\mu/2\pi = \SI{28}{\mega \hertz \per \milli \tesla}$.
The diamond is cut from a larger sample with an initial nitrogen concentration of approximately 200 ppm and naturally abundant $\!^{13}\mathrm{C}$ isotopes, treated with neutron irradiation with a total fluence of $\SI{5e17}{\per \centi \meter}$ for $\SI{50}{\hour}$ for the creation of lattice vacancies, and subsequently annealed at $\SI{900}{\celsius}$ for $\SI{3}{\hour}$ for the formation of \NV{} centers. The high amount of lattice damage due to the neutron irradiation and the high spin concentration are the reasons for the large inhomogeneous broadening $W/2\pi = \SI{8.65}{\mega \hertz}$ and short $T_2 = 1/\gamma_\perp = \SI{0.89}{\micro \second}$, see Supplementary Fig.~\ref{figSup:systemcharact}a,b. 

\vspace{0.4cm} \noindent \textbf{Hybrid system characterization} \vspace{0.2cm} \\ \noindent
We measure the inhomogeneously broadened spin distribution in the dispersive regime, as shown in Supplementary Fig.~\ref{figSup:systemcharact}a. This distribution is well described by a $q$-Gaussian function \cite{sandner2012strong}
\begin{equation}
    \rho(\Delta) = \frac{1}{\mathcal{C}}\left[1- \left(1-q \right)  \Delta^2/\delta_q^2\right]^{\frac{1}{1-q}} \,,
\end{equation}
with a normalization constant $\mathcal{C}$, $\delta_q = W \sqrt{ ( q-1 )/ (2^{q-1}-1)} $ and $W$ being the full width at half maximum, using a value for the shape parameter of $q=1.39$.
We determine the cavity linewidth $\kappa/2\pi = \SI{418}{\kilo \hertz}$, by measuring the steady-state transmission of the coupled system with a high input power of the vector network analyzer, see Supplementary Fig.~\ref{figSup:systemcharact}c. This brings the whole spin ensemble into a completely mixed state, effectively decoupling it from the cavity. Using low input power, we measure the normal-mode splitting of the system in the ground state. A steady-state solution of Eqs.~(\ref{eq:MBE_all}) without dipole-dipole interactions provides the fitted value of $g_\mathrm{coll}/2\pi = \SI{4.53}{\mega \hertz}$.
Comparing this value with the single spin coupling $g_0/2\pi \approx \SI{1.5}{\hertz}$ (obtained from a finite-element simulation of the cavity) we get an estimation of the number of spins $N = g_\mathrm{coll}^2/g_0^2 \approx \SI{9e12}{}$. Taking the volume of the diamond sample of around $\SI{5e6}{\cubic \micro \meter}$ and the carbon density in diamond of $n_c = \SI{1.755e23}{\per \cubic \centi \meter}$ into account results in an \NV{} concentration of approximately 10 ppm and a typical nearest-neighbor distance between \NVs{} of $r =(N/V)^{-1/3}\approx \SI{8}{\nano \meter}$.

\vspace{0.4cm} \noindent \textbf{Microwave setup, cavity, and inversion pulse} \vspace{0.2cm} \\ \noindent
We employ I/Q mixing with an arbitrary waveform generator and a microwave source at the cavity frequency to synthesize the inversion pulse. After amplification with a high-power amplifier, the pulse is injected into cavity port 1. The outgoing signal, exiting at cavity port 2, undergoes further amplification both inside and outside the cryostat. It is demodulated using another frequency source and recorded by a digitizer with a maximum sampling rate of $\SI{2}{\giga \hertz}$. Additional details on the superconducting cavity design and the protocol for initializing the inverted state, including details of the inversion pulse and detuning loop switching, can be found in Ref.~\cite{kersten2023triggered}.

\vspace{0.4cm} \noindent \textbf{Dipole-dipole interactions and spectral hole refilling} \vspace{0.2cm} \\ \noindent
To describe the refilling of depleted resonant spins, we consider the dipole-dipole interaction Hamiltonian,
\begin{equation}
    \mathcal{H}_\mathrm{dipole} = - \hbar  \sum_{j, k>j} \, { \frac{J_0}{|\mathbf{r}_{jk}|^3} }
    \frac{1}{\hbar^2}\left[ 3(\mathbf{S}_j\cdot \hat{\mathbf{u}}_{jk})(\mathbf{S}_k\cdot\hat{\mathbf{u}}_{jk}) - \mathbf{S}_j \cdot \mathbf{S}_k \right],
\end{equation}
where $J_0/2\pi= 51.9\,\mathrm{MHz\,nm^3}$, $\mathbf{r}_{jk}$ is the vector connecting the spins and $\hat{\mathbf{u}}_{jk}=\mathbf{r}_{jk}/|\mathbf{r}_{jk}|$. In our experiment, only the transition between the $m=0$ and $m=+1$ magnetic sublevels is of relevance. 
Taking into account the four different orientations of \NVs{} in our system, we can write the Hamiltonian as 
\begin{align}
    \mathcal{H}_\mathrm{dd} =\hbar\sum_{\substack{j, k\\ k>j}}\left[\left(J_{jk}\sigma_j^+\sigma_k^-+J_{jk}^*\sigma_k^+\sigma_j^-\right)+Q_{jk}\sigma_j^{ee}\sigma_k^{ee}\right],\\
    \mathrm{with}\quad J_{jk} = -\frac{J_0}{|r_{jk}|^3}(g_{jk}+\mathrm{i}h_{jk}),\quad Q_{jk}=-\frac{J_0}{|r_{jk}|^3},
    \label{eq:Jkl_Qkl}
\end{align}
where $g_{jk} = \tfrac{1}{2}(T_{jk}^{xx} + T_{jk}^{yy})$, $h_{jk} = \tfrac{1}{2}(T_{jk}^{xy} - T_{jk}^{yx}) $, and $q_{jk} = T_{jk}^{zz}$, with $T_{jk}^{\alpha\beta} = 3(\hat{\mathbf{e}}_{O_j}^\alpha\cdot\hat{\mathbf{u}}_{jk})(\hat{\mathbf{e}}_{O_k}^\beta\cdot\hat{\mathbf{u}}_{jk}) - \hat{\mathbf{e}}_{O_j}^\alpha\cdot \hat{\mathbf{e}}_{O_k}^\beta$. Here, $\hat{\mathbf{e}}_{O_j}^\alpha$ denotes the $\alpha$-oriented unit vector of the local coordinate system at site $j$, with $\alpha, \beta \in \lbrace x, y, z \rbrace$ ~(see Supplementary Information).

The total Hamiltonian is given by,
\begin{align}
    \mathcal{H} = &{\hbar} \sum_j \Delta_j \sigma^{ee}_j
    + \hbar g_0 \sum_j  \left(a^\dagger \sigma^-_j + \sigma_j^+ a \right) + \mathrm{i} \hbar \eta \left(a^\dagger - a \right) + \mathcal{H}_\mathrm{dd},
\end{align}
with $a^\dagger (a)$ being the creation (annihilation) operator of the cavity mode and $\sigma^{ee}_j, \sigma^\pm_j$ being the projection on the excited state and raising/lowering operators for the $j^{\mathrm{th}}$ spin, respectively. The cavity losses are accounted for by a rate $\kappa$. We disregard $T_1$ processes as the spin-lattice relaxation rate is negligible compared to all other dynamical timescales of our system. In our simulations, we also consider a weak external driving field $\eta$ to model the triggering of superradiant pulses by technical noise. It mainly affects the precise timing of the first revival pulse as well as its shape. 

We neglect three-spin and cavity induced contributions in the dynamics of the spin-spin coherences, $\expval*{\sigma_j^+\sigma_k^-}$, as they have negligible impact on the dynamics. Since the (single spin) decoherence rate $\gamma_\perp$ is the dominant timescale, we adiabatically eliminate the dynamics of these correlations, $\partial_t\expval*{\sigma_j^+\sigma_k^-}\approx 0$. Factorizing spin-cavity, $\expval*{a\sigma^+_j}\approx \expval*{a}\!\!\expval*{\sigma^+_j}$, and two-spin expectation values, $\expval*{\sigma_j^{ee}\sigma^-_k}\approx \expval*{\sigma_j^{ee}}\!\!\expval*{\sigma^-_k}$, we arrive at the dynamical equations of our system,
\begin{subequations}
\begin{align}
    &\partial_t\langle a\rangle = -\kappa\langle a\rangle -\mathrm{i} g_0 \sum_{j}\langle \sigma^-_j\rangle + \eta\, ,\label{eq:MBE_a}\\
    &\partial_t\langle\sigma_j^-\rangle = -(\gamma_\perp + \mathrm{i}\Delta_j) \langle\sigma^-_j\rangle + \mathrm{i} g_0 \langle a\rangle p_j\notag\\ &\,\qquad\qquad+\mathrm{i}p_j\sum_{\substack{k \\ k\neq j}} J_{jk}\expval*{\sigma_k^-}-\mathrm{i}\expval*{\sigma_j^-}\sum_{\substack{k \\ k\neq j}} Q_{jk}(p_k+1)/2\ ,\label{eq:MBE_sigma_m} \\
    &\partial_tp_j =
    -4g_0 \Im(\langle a^\dagger \rangle \langle\sigma^-_j\rangle) 
    -\sum_{\substack{k \\ k\neq j}} \frac{4\gamma_\perp |J_{jk}|^2\left(p_j-p_k\right)}{(\Delta_j-\Delta_k)^2+4\gamma_\perp^2} \, .\label{eq:MBE_sigma_z}
\end{align}\label{eq:MBE_all}
\end{subequations}
We sample $n_\mathrm{sim}=10^6$ \NV{} centers distributed randomly in a cube. For this number we find convergence for the cavity amplitude with respect to the spatial and frequency distribution as well as the orientations of the \NVs{}. In order to take into account the actual number $N$ of \NVs{} in the experiment, we have to consider ${N}/{n_\mathrm{sim}}$ copies of our box when coupling to the cavity in Eq.~\eqref{eq:MBE_a}. To avoid stiffness issues, we single out neighboring spins that equilibrate faster than all other time scales in the system and approximate them to instantaneously equilibrate. We simulate the remaining equations using an explicit Runge-Kutta method. A detailed derivation is provided in the Supplementary Information.

\vspace{0.4cm} \noindent \textbf{Superradiant instability threshold and cooperativity} \vspace{0.2cm} \\ \noindent
The instability criterion $p_0 C > 1$ is formally derived \cite{molmer1} from Eqs.~(\ref{eq:MBE_all}) as a necessary condition for the growth of the cavity amplitude starting from zero photons $\langle a \rangle = 0 $, in the case of uniform initial inversion $p_0 = \langle \sigma_j^z \rangle$ and vanishing $\langle \sigma_j^- \rangle = 0$, assuming no spin-spin interactions ($J_{jk} = Q_{jk} = 0$). 
The frequency-resolved single-spin cooperativity is given by
\begin{equation}
    C(\Delta) = \frac{g_0^2}{\kappa} \left( \gamma_\perp + \frac{\Delta^2}{\gamma_\perp} \right)^{-1} \,,
    \label{eq:CDelta}
\end{equation}
and defines the total cooperativity as an integral over the spin distribution (${\int}\rho(\Delta) \, d\Delta = N$):
\begin{equation}
    C = \int C(\Delta)\,\rho(\Delta)\,d\Delta = \frac{g_\mathrm{coll}^2}{\kappa \Gamma} \,,
    \label{eq:total_coop}
\end{equation}
where $\Gamma$ is the effective ensemble linewidth, incorporating both inhomogeneous broadening and intrinsic spin dephasing. For our system, we find $C = 15.2$ and $\Gamma/2\pi = \SI{3.22}{\mega \hertz}$.
For a non-uniform inversion profile $p(\Delta)$, this criterion generalizes in the continuum limit as a weighted integral over spin detunings. The system becomes unstable when
\begin{equation}
    \overline{pC} = \int p(\Delta)\, C(\Delta)\, \rho(\Delta)\, d\Delta > 1 \,.
    \label{eq:weighted_instability}
\end{equation}
This expression highlights that the instability is dominated by the spin inversion near resonance $p(\Delta\,{=}\,0)$, where $C(\Delta)$ is sharply peaked. Although this threshold strictly applies only in the absence of spin-spin interactions, it remains a valuable heuristic for interpreting the self-pulsing behavior.

Accordingly, the peak cavity amplitude during superradiant emission follows the relation 
\begin{equation} 
    \max(|a|) \propto p_0 - \frac{1}{C} \, 
    \label{eq:max_a_vs_p} 
\end{equation} 
as a consequence of the equations of motion Eqs.~(\ref{eq:MBE_all}). To motivate this relation, we consider the moment when the cavity field reaches its maximum, $\partial_t \langle a \rangle \approx 0$. Equation~(\ref{eq:MBE_a}) then implies $\langle a \rangle \propto \sum_j \langle \sigma_j^- \rangle \equiv S_-$. 
The transverse spin component $S_- = S_x -\mathrm{i}S_y$ is built up during the superradiant burst via the collective rotation of an initially $+z$-oriented ensemble and peaks simultaneously with the cavity amplitude in our superradiant regime. The system starts with $S_z \propto p_0 N$ and negligible initial coherence $S_- \approx 0$, the latter maintained at almost all times by rapid single-spin dephasing $\gamma_\perp$. Only during the superradiant emission does coherence build up dynamically.
This leads to a linear scaling $\max(|a|) \sim p_0 N$. The offset in Eq.~(\ref{eq:max_a_vs_p}) reflects the superradiant threshold: for $p_0 < 1/C$, the system remains stable and does not emit collectively. Since the dynamics are dominated by cavity-resonant spins, this scaling primarily reflects the on-resonance inversion $p(\Delta\,{=}\,0)$.

\begin{center}
\rule{4cm}{0.5pt}
\end{center}
\vspace{1cm}
\begin{center}
\textbf{\large Theory Supplement}
\end{center}

\vspace{0.4cm} \noindent \textbf{Dipole-dipole interactions for different \NV{} orientations} \vspace{0.2cm} \\ \noindent
In our microscopic description of the superradiant spin-cavity dynamics, we take into account the known form of the interaction between the magnetic dipoles of the \NV{} centers. Each dipole is oriented along one of the four directions defined by the diamond crystal, as illustrated in  Fig.~\ref{figSup:spin_directions}a. For each of the $N$ considered \NVs{}, we choose an axis $\hat{\mathbf{e}}^z_{O_k}$ that is aligned with one of these four directions, where the index $O_k$ labels the orientation, i.e., it takes on a random value out of $O_k\in\{1,2,3,4\}$ for each of the $k=1,\ldots,N$ nitrogen-vacancy centers. Correspondingly, we choose the remaining two cartesian axes $\hat{\mathbf{e}}^x_{O_k}$ and $\hat{\mathbf{e}}^y_{O_k}$ such that $\lbrace \hat{\mathbf{e}}^x_{O_k}, \hat{\mathbf{e}}^y_{O_k}, \hat{\mathbf{e}}^z_{O_k}\rbrace$ forms a right-handed orthogonal system. In this way, we can define the spin operators 
\begin{equation}
    \mathbf{S}_k = \hat{\mathbf{e}}^x_{O_k}s_k^x+\hat{\mathbf{e}}^y_{O_k}s_k^y+\hat{\mathbf{e}}^z_{O_k}s_k^z,
\end{equation}
in the local coordinate systems aligned with the individual axis of each \NV{}-center in terms of the corresponding spin-1 operators
\begin{align}
    s_k^x &= \frac{\hbar}{\sqrt{2}}(\lvert0\rangle_k\langle-\rvert \,+\, \lvert+\rangle_k\langle0\rvert \,+\, \lvert-\rangle_k\langle0\rvert+\lvert0\rangle_k\langle+\rvert),\\
    s_k^y &= \frac{\hbar}{\sqrt{2}{\rm i}}(\lvert0\rangle_k\langle-\rvert \,+\, \lvert+\rangle_k\langle0\rvert \,-\, \lvert-\rangle_k\langle0\rvert \,-\, \lvert0\rangle_k\langle+\rvert),\\
    s_k^z &= \hbar\left(\lvert+\rangle_k\langle+\rvert \,-\, \lvert-\rangle_k\langle-\rvert\right),
\end{align}
in the global laboratory frame, for which the $z$-axis is defined by the direction of the applied magnetic field which is chosen to coincide with the orientation of the diamond unit cell as shown in  Fig.~\ref{figSup:spin_directions}a. 
The energy eigenvalues of the \NVs{} are dominated by the zero-field splitting of about \SI{2.88}{\giga\hertz}, such that we can approximate their eigenvectors as those of the local $S^z_k$ operators. In addition, the applied magnetic field, ${\bf B}$, causes a linear Zeeman shift determined by the field projection $\hat{\mathbf{e}}_{O_k}^z\cdot \mathbf{B}$ onto the local $z$-axis of each \NV{} center. In our experiments, the magnetic field strength is adjusted such that the resulting energy of one of the $S^z_k$ sublevels is close to the cavity frequency of $3.1\,\mathrm{GHz}$. We define this level to be the $m=1$ state and can correspondingly neglect the $m=-1$ state, as its energy is far off resonance from the cavity. The relevant dynamics can thus be described in terms of effective spin-$1/2$ systems with respective transition and projection operators
\begin{align}\label{eq:spin12}
\sigma^-_k=&\lvert0\rangle_k\langle+\rvert,\\ 
\sigma^+_k=&\lvert+\rangle_k\langle0\rvert,\\ \sigma^{ee}_k=&\lvert+\rangle_k\langle+\rvert.
\end{align}
This is equivalent to choosing the local coordinate systems such that $\hat{\mathbf{e}}_{O_k}^z\cdot \mathbf{B}$ is positive for all four orientations $O_k$, the $m=1$ states are resonant with the cavity regardless of $O_k$. Specifically, this corresponds to (see Fig.~\ref{figSup:spin_directions}b)
\begin{equation}
    \hat{\mathbf{e}}^z_{O_k} \in \frac{1}{\sqrt{3}}\left\lbrace
    \left(
    \begin{matrix}
        1\\
        1\\
        1\\
    \end{matrix}
    \right),~
    \left(
    \begin{matrix}
        \phantom{-}1\\
        {-}1\\
        \phantom{-}1\\
    \end{matrix}
\right),~
    \left(
    \begin{matrix}
        -1\\
        \phantom{-}1\\
        \phantom{-}1\\
    \end{matrix}
\right),~
    \left(
    \begin{matrix}
        -1\\
        -1\\
        \phantom{-}1\\
    \end{matrix}
    \right)
    \right\rbrace.
    \label{eq:quantizationAxes}
\end{equation}
The remaining axes can then be constructed as cross products of $\hat{\mathbf{e}}^x_{O_k}$ and the global $z$-axis, $\hat{\mathbf{Z}}=(0,0,1)$, 
\begin{equation}
    \hat{\mathbf{e}}^x_{O_k} = \frac{\hat{\mathbf{Z}}\times\hat{\mathbf{e}}^z_{O_k}}{|\hat{\mathbf{Z}}\times\hat{\mathbf{e}}^z_{O_k}|},\quad\hat{\mathbf{e}}^y_{O_k}=\hat{\mathbf{e}}^z_{O_k}\times\hat{\mathbf{e}}^x_{O_k}.
    \label{eq:quantizationAxes2}
\end{equation}

\begin{extendedfigure}[t]
\includegraphics[width=75mm]{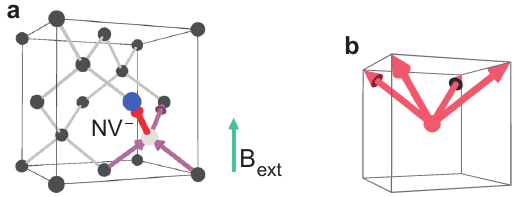}
\caption{
\textbf{Orientations of the \NV{} spins.}
\textbf{a,} Schematic of an \NV{} center within the diamond unit cell, showing one of the four possible crystallographic orientations defined by the axis connecting the substitutional nitrogen atom (blue) to the adjacent lattice vacancy (light gray).
\textbf{b,} Quantization axes used in the microscopic theory. As a result of the external magnetic field, the tetrahedral symmetry of the four \NV{} orientations is broken, and the axes form the edges of an inverted square pyramid. 
}
\label{figSup:spin_directions}
\end{extendedfigure}

The magnetic dipole-dipole interaction between the \NV{} centers is described by the Hamiltonian
\begin{equation}
    H_\mathrm{dd}=-\hbar \sum_{\substack{k,l\\ k\neq l}}\frac{\mu_0\gamma_\mathrm{e}^2\hbar}{4\pi |\mathbf{r}_{kl}|^3}\frac{1}{\hbar^2}\left[3(\mathbf{S}_k\cdot\hat{\mathbf{u}}_{kl})(\mathbf{S}_l\cdot\hat{\mathbf{u}}_{kl})-\mathbf{S}_k\cdot\mathbf{S}_l\right]\,,
\end{equation}
where, $\mu_0$ is the vacuum permeability, $\gamma_\mathrm{e}$ is the gyromagnetic ratio of the electron, and the unit vector $\hat{\mathbf{u}}_{kl}=\mathbf{r}_{kl}/r_{kl}$. We define the coupling constant 
\begin{equation}
    J_0 = \frac{\mu_0\gamma_\mathrm{e}^2\hbar}{4\pi} = 2\pi\times \frac{\mu_0 g_\mathrm{e}^2\mu_\mathrm{B}^2}{8\pi^2\hbar}=2\pi\times51.9\,\mathrm{MHz\,nm^3}.
\end{equation}
Substituting the local spin operators into the interaction Hamiltonian and neglecting counterrotating terms as well as transitions to the $m=-1$ level, as described above, yields the following effective Hamiltonian
\begin{align}
    H_\mathrm{dd}&=\hbar\sum_{\substack{k,l\\ k< l}}\left[\left(J_{kl}\sigma_k^+\sigma_l^-+J_{kl}^*\sigma_l^+\sigma_k^-\right)+Q_{kl}\sigma_k^{ee}\sigma_l^{ee}\right]
\end{align}
that describes the interaction between the spin-$1/2$ systems introduced in Eq.(\ref{eq:spin12}). 
The coupling constants 
\begin{equation}
    J_{kl} = -\frac{J_0}{|r_{kl}|^3}(g_{kl}+\mathrm{i}h_{kl}),\qquad Q_{kl} = -\frac{J_0}{|r_{kl}|^3}q_{kl}.
\end{equation}
are anisotropic with an angular dependence that is given by
\begin{align}
    &g_{kl} = \frac{1}{2}(T_{kl}^{xx} + T_{kl}^{yy}),\quad
    h_{kl} = \frac{1}{2}(T_{kl}^{xy} - T_{kl}^{yx}),\quad
    q_{kl} = T_{kl}^{zz},\notag\\
    &T_{kl}^{\alpha\beta} = 3(\hat{\mathbf{e}}_{O_k}^\alpha\cdot\hat{\mathbf{u}}_{kl})(\hat{\mathbf{e}}_{O_l}^\beta\cdot\hat{\mathbf{u}}_{kl}) - \hat{\mathbf{e}}_{O_k}^\alpha\cdot \hat{\mathbf{e}}_{O_l}^\beta,
    \label{eq:angDep}
\end{align} 
where \( \alpha, \beta \in \lbrace x, y, z \rbrace \).

\vspace{0.4cm} \noindent \textbf{Mean field dynamics} \vspace{0.2cm} \\ \noindent
The total Hamiltonian of our system can be written as
\begin{align}
    \mathcal{H} = &{\hbar} \sum_i \Delta_i \sigma^{ee}_i
    + \hbar g_0 \sum_i  \left(a^\dagger \sigma^-_i + \sigma_i^+ a \right) + i \hbar \eta \left(a^\dagger - a \right) + \mathcal{H}_\mathrm{dd},
\end{align}
with the spin-cavity detuning $\Delta$, the (individual) spin-cavity coupling strength $g_0$, cavity operators $a, a^\dagger$ and cavity drive $\eta$. In addition to the coherent dynamics by the Hamiltonian, we also consider the individual dephasing $\gamma_\perp$ as well as the cavity loss $\kappa$ in a Lindblad master equation, while we neglect $T_1$ processes. The relevant equations read
\begin{align}
    &\partial_t\langle a\rangle = -\kappa\langle a\rangle -\mathrm{i} g_0 \sum_{j}\langle \sigma^-_j\rangle + \eta ,\label{eq:supMBE_a}\\
    &\partial_t\langle\sigma_j^-\rangle = -(\gamma_\perp + \mathrm{i}\Delta_j) \langle\sigma^-_j\rangle + \mathrm{i} g_0 \langle a \sigma^{z}_j\rangle +\mathrm{i}\sum_{\substack{k \\ k\neq j}} J_{jk}\expval*{\sigma^z_j\sigma_k^-}-\mathrm{i}\sum_{\substack{k \\ k\neq j}} Q_{jk}\langle \sigma_j^-\sigma^{ee}_k\rangle \,,\label{eq:supMBE_sigma_m} \\
    &\partial_t\langle\sigma_i^+\sigma_j^-\rangle = -[\mathrm{i}(\Delta_j-\Delta_i)+2\gamma_\perp]\langle\sigma_i^+\sigma_j^-\rangle + \mathrm{i}g_0(\expval*{a\sigma_i^+\sigma_j^z}-\expval*{a^\dagger\sigma_i^z\sigma_j^-})-\mathrm{i} J_{ji}(\expval*{\sigma^{ee}_i}-\expval*{\sigma^{ee}_j})
    \notag\\ 
    &\qquad\qquad\qquad +\mathrm{i}\sum_{\substack{l\\l\neq i,j}}(J_{jl}\expval*{\sigma^+_i\sigma^z_j\sigma_l^-}-J_{il}^*\expval{\sigma_l^+\sigma_i^z\sigma_j^-})
    -\mathrm{i}\sum_{\substack{k\\k\neq i,j}}(Q_{kj}\expval*{\sigma^+_i\sigma_k^{ee}\sigma_j^-}-Q_{ki}\expval*{\sigma^+_i\sigma_k^{ee}\sigma_j^-})\label{eq:coherences}\\
    &\partial_tn_j =
    -2g_0 \Im(\langle a^\dagger \sigma^-_j\rangle) 
    -\mathrm{i}\sum_{\substack{k \\ k\neq j}}\left(J_{jk}\langle\sigma_j^+\sigma_k^-\rangle- J_{kj}\langle\sigma_k^+\sigma_j^-\rangle\right) \, ,\label{eq:density_beforeAdEl}
\end{align}
where we defined $n_j=\expval{\sigma^{ee}_j}$.
The second line of Eq.~\eqref{eq:coherences} contains terms that involve products of three spin operators. We numerically exclude their relevance on the refilling dynamics and neglect them in the following. We further neglect the cavity contributions to this equation because their effect on the refilling is suppressed with $g_0^2/\kappa$. Since $\gamma_\perp$ is the dominant timescale in the dynamics, we can now adiabatically eliminate $\langle\sigma_i^+\sigma_j^-\rangle$ and reinsert the expression into Eq.~\eqref{eq:density_beforeAdEl}. Finally, we factorize spin-cavity, $\expval*{a\sigma^+_j}\approx \expval*{a}\!\!\expval*{\sigma^+_j}$, and two-spin expectation values, $\expval*{\sigma_j^{ee}\sigma^-_i}\approx \expval*{\sigma_j^{ee}}\!\!\expval*{\sigma^-_i}$ and arrive at the dynamical equations
\begin{subequations}
\begin{align}
    &\partial_t\langle a\rangle = -\kappa\langle a\rangle -\mathrm{i} g_0 \sum_{j}\langle \sigma^-_j\rangle + \eta ,\label{eq:supMBE_a_2}\\
    &\partial_t\langle\sigma_j^-\rangle = -(\gamma_\perp + \mathrm{i}\Delta_j) \langle\sigma^-_j\rangle + \mathrm{i} g_0 \langle a\rangle (2n_j-1)+\mathrm{i}(2n_j-1)\sum_{\substack{k \\ k\neq j}} J_{jk}\expval*{\sigma_k^-}-\mathrm{i}\expval*{\sigma_j^-}\sum_{\substack{k \\ k\neq j}} Q_{jk}n_k\ ,\label{eq:supMBE_sigma_m_1} \\
    &\partial_tn_j =
    -2g_0 \Im(\langle a^\dagger \rangle \langle\sigma^-_j\rangle) 
    -\sum_{\substack{k \\ k\neq j}} \frac{4\gamma_\perp |J_{jk}|^2}{(\Delta_j-\Delta_k)^2+4\gamma_\perp^2} \left(n_j-n_k\right)\, .\label{eq:supMBE_sigma_z}
\end{align}
\end{subequations}

\vspace{0.4cm} \noindent \textbf{Numerical simulation} \vspace{0.2cm} \\ \noindent
In our simulations, we consider up to $n_\mathrm{sim}=10^6$ spins, thereby ensuring statistical significance when drawing from the distributions. We sample the spin frequencies from the experimentally determined distribution and consider the spins randomly distributed in a cube of appropriate size. We further randomly sample the orientations of the \NV{} centers. To correctly capture the spin-cavity interaction, we consider $N/n_\mathrm{sim}$ copies of the cube in the relevant sums of the above equations. 
We simulate these equations using an explicit Runge-Kutta method which is suitable to the spin-cavity dynamics. The dipole-dipole induced relaxation process introduces stiffness issues in the simulation. We tackle these issues by singling out strongly interacting neighbors and treating them as instantaneously relaxed, and project the equations on the corresponding reduced subspace. 

\vspace{0.4cm} \noindent \textbf{Qualitative solution to the refilling dynamics} \vspace{0.2cm} \\ \noindent
While the full numerical solution is available,
it is instructive to consider an approximated scenario using the so-called \textit{relaxation time approximation} to derive an analytical solution to the refilling dynamics. Note that the rates we derive within the approximation are far beyond the rates observed experimentally. Still, qualitative insights applicable to the scenario in the main text may be extracted from the model. 
We focus on the case of refilling in the absence of the cavity. 
Within the relaxation time approximation, we approximate that for each depopulated spin $i$ all its neighbors $k$ have already reached equilibrium, $n_k\approx n_\mathrm{avg}$, yielding the dynamical equations
\begin{equation}
    \partial_t \Delta n_i = -\sum_{\substack{k\\ k\neq i}}\frac{4\gamma |J_{ki}|^2}{(\Delta_k-\Delta_i)^2+4\gamma^2}\Delta n_i,
\end{equation}
where we defined $\Delta n_i=n_i-n_\mathrm{avg}$. This decouples the equations for different $\Delta n_i$, such that we can write the solution
\begin{equation}
    \Delta n_i(t) = \exp\left(-\sum_{\substack{k\\ k\neq i}}\frac{4\gamma |J_{ki}|^2 t}{(\Delta_k-\Delta_i)^2+4\gamma^2}\right)\Delta n_i(0).
\end{equation}
We can now perform the ensemble average on this solution regarding the detuning $\Delta$, position $\vec{r}$ and the orientations~$O$, 
\begin{align}
    &\expval{\exp\left(-\sum_{\substack{k\\ k\neq i}}\frac{4\gamma |J_{ki}|^2 t}{(\Delta_k-\Delta_i)^2+4\gamma^2}\right)} =\qquad\qquad\notag\\ &=\int\dd{\Delta_1}...\dd{\Delta_{N-1}}\bar{n}(\Delta_1)...\bar{n}(\Delta_{N-1})\int\frac{\dd{\vec{r}_1}}{V}...\frac{\dd{\vec{r}_{N-1}}}{V}\frac{1}{4}\sum_{O_1}...\frac{1}{4}\sum_{O_{N-1}}\times\notag\\
    &\qquad\qquad\qquad\qquad\times\exp\left(-\sum_{\substack{k\\ k\neq i}}\frac{4\gamma |J_{ki}|^2 t}{(\Delta_k-\Delta_i)^2+4\gamma^2}\right)=\notag\\ &=\left[\int\dd{\Delta_k}\bar{n}(\Delta_k)\int\frac{\dd{\vec{r}}}{V}\frac{1}{4}\sum_{O_k}\exp\left(-\frac{4\gamma  t[g_{ki}^2+h_{ki}^2]}{(\Delta_k-\Delta_i)^2+4\gamma^2}\frac{J_0^2}{r^6}\right)\right]^{N-1}\notag\\ &=\left[1-\frac{1}{N-1}\int\dd{\Delta_k}\bar{n}(\Delta_k)\frac{1}{4}\sum_{O_k}\frac{N-1}{V}\int {\dd{\vec{r}}}\left[1-\exp\left(-\frac{4\gamma  t[g_{ki}^2+h_{ki}^2]}{(\Delta_k-\Delta_i)^2+4\gamma^2}\frac{J_0^2}{r^6}\right)\right]\right]^{N-1}\notag\\
    &\xrightarrow{N\rightarrow\infty}\exp\left(\int\dd{\Delta_k}\bar{n}(\Delta_k)\frac{N}{V}\frac{1}{4}\sum_{O_k}\int\sin(\theta)\dd{\theta}\dd{\varphi}\int {r^2\dd{{r}}}\times\right.\notag\\&\left.\qquad\qquad\qquad\qquad\times\left[1-\exp\left(-\frac{4\gamma  t[g_{ki}^2+h_{ki}^2]}{(\Delta_k-\Delta_i)^2+4\gamma^2}\frac{J_0^2}{r^6}\right)\right]\right)\notag\\
    &=\exp\left(\frac{N}{V}\frac{4\pi^{\frac{3}{2}}}{3}\int\dd{\Delta_k}\bar{n}(\Delta_k)\frac{1}{4}\sum_{O_k}\frac{1}{4\pi}\int\sin(\theta)\dd{\theta}\dd{\varphi}\sqrt{\frac{4\gamma  t{J_0^2}[g_{ki}^2+h_{ki}^2]}{(\Delta_k-\Delta_i)^2+4\gamma^2}}\right) \notag\\ &=: \exp(-\sqrt{t/{T_\mathrm{r}}(\Delta_i)}),
\end{align}
where we defined
\begin{align}
    {T}_\mathrm{r}^{-1}(\Delta) &:= \left(\frac{N}{V}\right)^2\frac{J_0^2}{\gamma}\frac{16\pi^3}{9}\nu^2(\Delta)\xi^2(O),\\ & \quad\nu(\Delta)=\int\dd{\Delta'}\bar{n}(\Delta')\sqrt{\frac{4\gamma^2 }{(\Delta'-\Delta)^2+4\gamma^2}},\\ &\quad \xi(O) = \frac{1}{4}\sum_{O'}\zeta_{OO'},\quad\zeta_{OO'}=\frac{1}{4\pi}\int\sin(\theta)\dd{\theta}\dd{\varphi}\sqrt{g_{OO'}^2+h_{OO'}^2}.\label{eq:angAVG}
\end{align}
Note that we write $g_{OO'}$ and $h_{OO'}$ with orientation indices, since they only depend on the orientation and the angle, but not on the spin-spin distance. The angular average is parametrized in spherical coordinates, i.e., $\hat{\mathbf{u}}$ is replaced by the unit vector in spherical coordinates in Eq.~\eqref{eq:angDep} before the average. Different orientations $O$ sample the corresponding coordinate systems from Eqs.~\eqref{eq:quantizationAxes} and \eqref{eq:quantizationAxes2}. For the angular average over different \NV{} orientations we find 
\begin{equation}
    (\zeta_{OO'}) = \left(
    \begin{matrix}
        0.38&0.65&0.65&0.83\\
        0.65&0.38&0.83&0.65\\
        0.65&0.83&0.38&0.65\\
        0.83&0.65&0.65&0.38\\
    \end{matrix}
    \right). \label{eq:subensembleAngAvg}
\end{equation}
Note that it was already observed in \cite{choi2017depolarization} that on average the relaxation of spins within one subensemble of a specific orientation is lower than the inter-subensemble relaxation. For our specific setup with 4 subensembles on resonance, we find that there is a further distinction between different inter-subensemble relaxation rates. Namely, for each orientation, there exist two different orientations with relative relaxation weight 0.65 and one with relative relaxation weight 0.83 [cf. Eq.~\eqref{eq:subensembleAngAvg}]. This can be understood when considering that for a chosen quantization axis in Fig.~\ref{figSup:spin_directions}b, two other (directed) quantization axes enclose an angle of 70.5° with that axis while the third axis encloses an angle of 109.5°. 
That way, we find $\xi^2=0.397$ for the angular average. The frequency average is determined from the experimental spin distribution; on resonance, we arrive at $\nu^2(\Delta=0)=0.0480$. Using the parameters of Fig.~3 in the main text, we thus find ${T_\mathrm{r}}(\Delta=0)=1.14\mathrm{\upmu s}$. While the on-resonance relaxation of the inversion also follows a stretched exponential, the experiment yields a considerably longer relaxation timescale of about ${T_\mathrm{r}}=11.6\,\mathrm{\upmu s}$ (see Fig.~\ref{fig2:secondholdtime} in the main text). Note that although the timescales are quite different, we find that some features of the relaxation persist qualitatively in the full simulation. First, the on-resonance population relaxes with a stretched exponential behavior. Second, we find the on-resonance relaxation rate to be the same for different initial inversions $p_0$ before the first superradiant decay, and thus different shapes of the spectral hole. This is evident in the approximate solution, where the different frequencies relax completely independently of each other and is consistent with the observations in Supplementary Fig.~\ref{figSup:hole_depth}. 

\end{document}